\documentclass[12pt]{article}
\usepackage[body={18cm, 23cm, centered}]{geometry}
\usepackage{amsmath,amssymb,amsthm,amscd,cite,comment,amsfonts,indentfirst,color,setspace,bbold,arydshln}
\usepackage{mathtext,marvosym,textcomp}
\usepackage[makeroom]{cancel}
\usepackage{mathtools}
\usepackage{braket}
\usepackage{cancel}
\usepackage{float}

\usepackage[debug,pageanchor=false]{hyperref}
\hypersetup{colorlinks=true,linktocpage,breaklinks,
	urlcolor=blue,
	linkcolor=blue,
	citecolor=blue
}

\usepackage[vcentermath]{youngtab}

\usepackage[mathmode,centertableaux]{ytableau}
\usepackage{tikz}
\usetikzlibrary{arrows}
\usetikzlibrary{shapes.misc}
\usepackage{tikz-cd}

\usetikzlibrary{arrows,shapes,positioning, fit}
\tikzstyle{every picture}+=[remember picture]
\tikzstyle{na} = [baseline=-.5ex]
\tikzstyle{format} = [rounded rectangle,
                      thick,
                      minimum size=1cm,
                      draw=red!00!black!80,
                      top color=white,
                      bottom color=red!00!black!00,
                      on grid]
\tikzstyle{format1} = [rectangle,
                      thick,
                      minimum size=1cm,
                      draw=red!00!black!80,
                      top color=white,
                      bottom color=red!00!black!00,
                      on grid]
\tikzstyle{format0} = [rounded rectangle,
					  thick,
				      minimum size=1.2cm,
					  draw=blue!00!black!80,
					  top color=white,
                      bottom color=blue!00!black!00,
                      text centered]
\tikzstyle{formatd} = [rounded rectangle,
					  thick,
					  dashed,
				      minimum size=1cm,
					  draw=blue!50!black!80,
					  top color=white,
                      bottom color=blue!00!black!00,
                      text centered]
\tikzstyle{format1d} = [rounded rectangle,
                      thick,
                      dashed,
                      minimum size=2cm,
					  draw=blue!50!black!80,
					  top color=white,
                      bottom color=blue!00!black!00,                  
                      text centered]
                      
\tikzset{cross/.style={cross out, draw=black, minimum size=2*(#1-\pgflinewidth), inner sep=0pt, outer sep=0pt},
	cross/.default={5pt}}
\usepackage[ normalem]{ulem}
\usepackage{tocloft}

\numberwithin{equation}{section}

\selectfont

\def\a{\alpha} 
\def\b{\beta} 
 
\def\d{\delta} 
\def\e{\epsilon}
\def\ve{\varepsilon}

\def\k{\kappa} 
 
\def\m{\mu}
\def\n{\nu} 
\def\p{\pi}
\def\x{\xi} 
\def\r{\rho}

\def\s{\sigma} 
  
\def\f{\phi}
\def\y{\psi}

\def\F{\Phi}

\def\W{\Omega}

\def\fr{\frac}  \def\dt{\partial}

\def\mc{\mathcal}

\def\XX{\mathbb{X}}
\def\RR{\mathbb{R}}
\def\SS{\mathbb{S}}

\def\Tr{\mathrm{Tr}}

\def\rmGL{\mathrm{GL}}
\def\rmSO{\mathrm{SO}}
\def\rmSL{\mathrm{SL}}
\def\rmSU{\mathrm{SU}}
\def\rmU{\mathrm{U}}

\def\rmE{\mathrm{E}}

\def\frso{\mathfrak{so}}
\def\frsu{\mathfrak{su}}
\def\frsl{\mathfrak{sl}}
\def\frgl{\mathfrak{gl}}
\def\frg{\mathfrak{g}}

\begin{document}
\renewcommand{\refname}{\begin{center}References\end{center}}
	
\begin{titlepage}
		
	\vfill
	\begin{flushright}

	\end{flushright}
		
	\vfill
	
	\begin{center}
		\baselineskip=16pt
		{\Large \bf 
		Tri-vector deformations on compact isometries
		}
		\vskip 1cm
            Edvard T. Musaev$^{a,b}$\footnote{\tt musaev.et@phystech.edu}, 
		    Timophey Petrov$^{a}$\footnote{\tt petrov.ta@phystech.edu},
		\vskip .3cm
		\begin{small}
			{\it 
				$^a$Moscow Institute of Physics and Technology,
			    Institutskii per. 9, Dolgoprudny, 141700, Russia,\\
				$^b$Kazan Federal University, Institute of Physics, Kremlevskaya 16a, Kazan, 420111, Russia\\
			}
		\end{small}
	\end{center}
		
	\vfill 
	\begin{center} 
		\textbf{Abstract}
	\end{center} 
	\begin{quote}
         Classical Yang-Baxter equation governing bi-vector deformations of 10d supergravity is known to have no solutions along non-abelian compact isometries. By providing explicit examples we show that this is in contrast to generalized Yang-Baxter equation governing tri-vector deformations of 11d supergravity. We present deformations of the AdS${}_7\times \mathbb{S}^4$ and flat backgrounds with isometries generated by Killing vectors of a sphere. Isometries of the AdS space-time are preserved by such deformations.
	\end{quote} 
	\vfill
	\setcounter{footnote}{0}
\end{titlepage}
	
\clearpage
\setcounter{page}{2}
	
\tableofcontents

\section{Introduction}

When studying a ((super)conformal) quantum field theory one is interested whether it belongs to a family of theories connected by an RG flow or is an isolated point in the moduli space. An RG flow from a UV fixed point can be triggered by adding a(n ir)relevant operator, or by turning on VEV for an operator (see e.g. \cite{deBoer:1999tgo,Skenderis:2002wp}).  An example of such irrelevant operator would be $\f^4$ in the 4d theory of a real scalar field, that flows to a free theory in the IR. Note that this operator is classically marginal and quantum irrelevant. A closely related but different question is the study of manifolds of fixed points generated by marginal operators in a given CFT \cite{Strassler:1998iz,Cordova:2016xhm,Bashmakov:2017rko}. In contrast to a(n ir)relevant operator an exactly marginal operator does not trigger an RG flow changing a UV theory to a different theory in IR. Instead the theory stands in the same fixed point, that could be a fixed line or a fixed manifold. An example of a family of (exactly) marginal operators generating a manifold of fixed point is the Leigh-Strassler deformation of $d=4$ $\mc{N}=4$ super Yang-Mills theory preserving $\mc{N}=1$ supersymmetry \cite{Leigh:1995ep}. This two-parameter deformation labeled by $\b$ and $\rho$ can be described in terms of the following superpotential
\begin{equation}
    \label{eq:W}
    W = \Tr \Big[\big(e^{i\b}\F_1\F_2\F_3 - e^{-i\b}\F_1\F_3\F_2\big) + \r \big(\F_1^2 + \F_2^2 +\F_3^2\big)\Big].
\end{equation}
From the point of view of AdS/CFT correspondence the interpretation of the $\r$-deformation is not clear\footnote{Although some progress has been made recently \cite{Ashmore:2021mao}} while the supergravity solution dual to the $\beta$-deformed SYM theory has been presented in \cite{Lunin:2005jy}. In the language relevant to the results of the present paper this is a Yang-Baxter deformation of AdS$_5\times \SS^5$ background defined by a bivector $\b = \dt_{\phi_1}\wedge \dt_{\phi_2}$, where $\phi_{1,2}$ are coordinates along two isometric circles inside the 5-sphere. Such deformation breaks the internal $\rmSU(4)$ isometry  down to $\rmU(1)$ and hence breaks $\mc{N}=4$ supersymmetry to $\mc{N}=1$. For the metric and the B-field the deformed background $G,B$ is related to the initial metric $g$ and the deformation bi-vector by
\begin{equation}
    \label{eq:openclosed}
    G+B = (g^{-1} + \b)^{-1}.
\end{equation}
A generalization of this approach to arbitrary backgrounds with a set of at least two Killing vectors $k_a$ has been developed in \cite{Matsumoto:2014nra,Araujo:2017jkb,Araujo:2017jap,Bakhmatov:2017joy, Bakhmatov:2018apn,Bakhmatov:2018bvp}, where the bi-vector parameter has been taken in the form
\begin{equation}
    \b = r^{ab}k_a \wedge k_b,
\end{equation}
with a constant antisymmetric matrix $r_{ab} = - r_{ba}$. It has been shown that in order for the map \eqref{eq:openclosed} (together with similar maps for the RR sector) to generate a supergravity solution it is enough that the matrix $r_{ab}$ satisfies classical Yang-Baxter equation
\begin{equation}
    r^{[a_1 |b_1|}r^{a_2|b_2|}f_{b_1b_2}{}^{a_3]}=0
\end{equation}
and the so-called unimodularity condition $r^{ab}f_{ab}{}^c=0$ introduced in \cite{Borsato:2016ose}. Here $f_{ab}{}^c$ denote structure constants of the algebra of Killing vectors $[k_a,k_b] = f_{ab}{}^c k_c$, which is non-abelian in the most general case. It is worth to mention, that the notion of Yang-Baxter deformations has appeared earlier in the context of deformations of 2d $\s$-models that preserve integrability \cite{Klimcik:2002zj,Klimcik:2008eq,Delduc:2013qra} (for a review see \cite{Thompson:2019ipl,Seibold:2020ouf}). Assuming that a background on which the string $\s$-model is integrable, is dual to an integrable gauge theory, this allows to speculate that Yang-Baxter deformations generate families of integrable CFT's. To our knowledge no examples directly supporting such a statement are known in the literature.

In the context of AdS/CFT correspondence, i.e. when the initial background has an AdS factor, it is natural to consider three separate cases, depending on whether all Killing vectors are taken along i) isometries of the AdS space, ii) isometries of the internal space (say a sphere), or iii) are mixed. On top of that a deformation can be abelian if all isometries commute, or non-abelian if not. In the case of abelian isometries the meaning of all three classes of YB deformations on the gauge theory side is clear \cite{Lunin:2005jy,Imeroni:2008cr}: the deformation replaces products of fields by star product defined by the action of the isometry generators. For a deformation along two generators $Q_{1,2}$ schematically this can be written as
\begin{equation}
    f\,g \to f*g = \exp\Big[\b (Q_1^f Q_2^g - Q_2^f Q_1^g)\Big]f\,g,
\end{equation}
where the superscript indicates on which field a generator acts. Here we see the crucial difference between deformations along external or mixed isometries and those along internal isometries. With  both generators taken along the AdS space the above formula reproduces the well-known Moyal star product rendering the deformed theory to be a non-commutative gauge theory. The mixed case corresponds to the so-called dipole deformation and eventually is also adding non-locality to the theory \cite{Bergman:2000cw,Bergman:2001rw}. Finally, taking both isometries along the internal symmetry group we end up with an expression like \eqref{eq:W}, where fields simply acquire additional factors depending on their charges w.r.t. the isometry generators. This is precisely what happens in the case of the  Lunin-Maldacena deformation dual to the $\b$-deformation of Leigh-Strassler. We conclude that only deformations along isometries of the internal compact space can be interpreted as adding operators in the theory. Further extension of the above logic for this kind of deformations to non-abelian isometries is prevented by a Theorem stating that for compact groups classical Yang-Baxter equation implies either $r_{ab}=0$ or $f_{ab}{}^c = 0$ \cite{Lichnerowicz:1988abc, Pop:2007abc} (see Appendix \ref{app:th} for a brief review). Hence, along compact isometries only abelian bi-vector Yang-Baxter deformations are allowed.

Recently a generalization of bi-vector Yang-Baxter deformations to backgrounds of 11d supergravity has been proposed \cite{Bakhmatov:2019dow,Bakhmatov:2020kul,Gubarev:2020ydf}. These are generated by a tri- and a six-vectors that is dictated by the structure of the corresponding U-duality groups \cite{Sakatani:2019zrs,Malek:2019xrf,Malek:2020hpo}. Explicit expressions are more complicated than \eqref{eq:openclosed} and for the SL(5) case convenient for backgrounds of the form $M_7\times M_4$ we provide below in \eqref{Cdef}. The tri-vector parameter $\W$ is again taken in the poly-Killing ansatz
\begin{equation}
    \W = \r^{abc} k_a\wedge k_b \wedge k_c,
\end{equation}
and a sufficient set of conditions for such deformation to generate a solution to supergravity equations is 
\begin{equation}
    \label{eq:gencybe}
    \begin{aligned}
        \rho^{i_1i_2i_3}f_{i_2i_3}^{\ \ \ \, i_4}=0,&&\\
        \rho^{i_1[i_2|i_6|}\rho^{i_3i_4|i_5|}f_{i_5i_6}^{\ \ \ \, |i_7]}-\rho^{i_2[i_1|i_6|}\rho^{i_3i_4|i_5|}f_{i_5i_6}^{\ \ \ \, |i_7]}=0.
    \end{aligned}
\end{equation}
The first line is an analogue of the unimodularity condition, and the second line is usually referred to as the generalized Yang-Baxter equation. The very first abelian example of such tri-vector deformations has been presented in \cite{Lunin:2005jy} although from different premises. See also \cite{Ashmore:2018npi} for examples of such deformations from generalized geometry. Since tri-vector deformations can also be defined for type IIA/B supergravity backgrounds one becomes interested in the same question: what gauge theory is dual to a given tri-vector deformation? Given the results of \cite{Imeroni:2008cr} of particular interest are deformations along compact isometries as these presumably correspond to adding (exactly marginal) operators. Probably via somehow defined star tri-product (see e.g. \cite{Gustavsson:2010nc} and the review \cite{Gubarev:2023jtp}). The Theorem that prevents non-abelian bi-vector Yang-Baxter deformations along compact isometries  cannot immediately be  applied to tri-vector as it essentially involves matrices with two indices. 

In this work we search for solutions to the generalized classical Yang-Baxter equation \eqref{eq:gencybe} for the Lie algebra of the compact group $\rmSO(5)$. We find several such solutions and construct the corresponding tri-vector deformations of the AdS$_7\times \SS^4$ and flat backgrounds, that solve equations of 11d supergravity. In addition we find a large set of tri-vector deformations that satisfy the first line of \eqref{eq:gencybe} and do not satisfy the second line. In the special case of deformations of 4d manifolds, that we focus here on, such deformations are still solutions to supergravity equations. Since the AdS space stays intact our solutions in principle can be understood as dual to (exactly) marginal deformations of $D=6$ $\mc{N}=(2,0)$ theory dual to AdS${}_7\times \SS^4$. The backgrounds we find preserve no supersymmetry.

Structure of the paper is the following. In Section \ref{sec:exft} we briefly describe the relevant parts of the formalism of the SL(5) exceptional field theory that provides a convenient packing of supergravity fields for defining tri-vector deformations. In Section \ref{sec:sols} we provide explicit solutions to generalized Yang-Baxter equation in terms of abstract $\frso(5)$ generators in Cartan basis, and construct their realization in terms of Killing vectors on a sphere. In Section \ref{sec:ex} we construct explicit examples of tri-vector deformations of the AdS$_7\times \SS^4$ and flat backgrounds. Section \ref{sec:conc} is devoted to discussion of the results, their relevance to problems of gauge-gravity duality and possible further extensions. In Appendix \ref{app:th} we give details of the Theorem stating that for compact groups only abelian solutions of the classical Yang-Baxter equation are allowed.

\section{Tri-vector deformations}
\label{sec:exft}

Tri-vector generalized Yang-Baxter deformations of 11d supergravity backgrounds have been observed in \cite{Bakhmatov:2019dow,Bakhmatov:2020kul,Gubarev:2020ydf} as special transformations within the group of local U-duality transformations of exceptional field theory. In the special case of group manifolds such deformations generate a family of exceptional Drinfeld algebras \cite{Sakatani:2019zrs,Malek:2019xrf}. To setup notations and list expressions for further reference let us briefly review the formalism. In this work we consider tri-vector deformations within the SL(5) U-duality group, that given the truncation of \cite{Bakhmatov:2019dow,Bakhmatov:2020kul} restricts us to backgrounds of the form $M_7\times M_4$, where $M_7$ is a manifold of constant curvature. Exceptional field theory is a U-duality covariant formulation of both 11d and Type IIB supergravities, whose field content in the SL(5) case is the following
\begin{equation}
    \begin{aligned}
        & g_{\m\n}, && A_\m{}^{MN}, && B_{\m\n M}, && m_{MN}.
    \end{aligned}
\end{equation}
Here and in what follows we adopt the following index conventions
\begin{equation}
    \begin{aligned}
        & \m,\n,\dots =0,1,\dots,6 && \text{external curved indices}, \\
        & m, n, k, l, \dots = 1,2,3,4, && \text{internal curved indices},\\
        & M,N,K, \dots = 1,\dots,5 && \text{indices labeling the }\bf{5}\text{ of SL(5)}, \\
        & a,b,c,d = 1,\dots, N && \text{indices labeling Killing vectors.}
    \end{aligned}
\end{equation}
Hence, the fields of the theory include the 7d metric, ten vector fields, five 2-forms and 14 scalar fields packed into the coset representative
\begin{equation}
    m_{MN} \in \fr{\rmSL(5)}{\rmSO(5)}.
\end{equation}
In general fields of exceptional field theory live on a special extended space labeled by 10 coordinates  $\XX^{MN}$ and satisfy a condition called the section constraint, that effectively restricts this dependence and is needed for consistency of local symmetries of the theory (for more details see the reviews \cite{Baguet:2015xha,Hohm:2019bba,Musaev:2019zcr,Berman:2020tqn} and original papers \cite{Hohm:2013pua,Musaev:2015ces}). Here we assume that all fields depend only on coordinates of the normal space-time $(x^\m,x^m)$, hence for us equations of exceptional field theory are simply a more convenient form of equations of the conventional 11d supergravity.

We set $A_\m{}^{MN} = 0$ and $B_{\m\n M} = 0$ following \cite{Bakhmatov:2019dow,Bakhmatov:2020kul} where it has been shown that such truncated equations of exceptional field theory do not generate these gauge fields again. Since we consider backgrounds of the form $M_7\times M_4$ we further assume that dependence of the external metric $g_{\m\n}$ on the coordinates $(x^\m,x^m)$ factorizes as follows
\begin{equation}
    g_{\m\n} = e^{-2\f(x^m)} g^{\fr15}(x^m) \bar{g}_{\m\n}(x^\m),
\end{equation}
where $g=\det g_{mn}$ is the determinant of the metric on $M_4$. Hence, full eleven-dimensional interval reads
\begin{equation}
    ds^2 = e^{-2\f(x^m)}  \bar{g}_{\m\n}(x^\m) dx^\m dx^\n + g_{mn}(x^m)dx^m dx^n.
\end{equation}
The metric $g_{mn}$ and the 3-form $c_{mnk}$ parameterize the coset
\begin{equation}
    M_{MN} = e^{\f}
        \begin{bmatrix}
                g^{-\fr12} g_{mn} & - v_n\\
                v_m & g^{\fr12}(1+v^2)
        \end{bmatrix} \in \fr{\rmGL(5)}{\rmSO(5)}
\end{equation}
related to $m_{MN}$ by a rescaling $m_{MN} = e^{-\f} g^{\fr1{10}} M_{MN}$. Here we define the vector
\begin{equation}
    v^m = \fr1{3!} \ve^{mnkl}c_{nkl},
\end{equation}
assuming it only depends on the coordinates $x^m$ and  use $\ve^{mnkl} = g^{-\fr12}\e^{mnkl}$ as the totally antisymmetric tensor. The introduced rescaling and factorization of coordinate dependence are needed to define tri-vector deformation in the truncation ansatz, that is a special SL(5) transformation acting linearly at $M_{MN}$ and $\bar{g}_{\m\n}$. The latter is simply a singlet as well as $g_{\m\n}$, while  the former transforms as
\begin{equation}
    \begin{aligned}
        M_{MN} & \to O_M{}^K O_N{}^L M_{KL}, \\
        O_M{}^N & = \exp \big[\Omega^{mnk}T_{mnk}\big]{}_M{}^N \in \rmSL(5).
    \end{aligned}
\end{equation}
The generators $(T_{mnk})_M{}^N$ are negative level generators of SL(5) w.r.t. to the GL(1) factor in the decomposition $\rmSL(5)\hookleftarrow \rmGL(1) \times \rmSL(4)$. To be more precise, generators $T_M{}^N$ of the algebra $\frsl(5)$ in the fundamental representation read
\begin{equation}
    (T_M{}^N)_K{}^L = \d_M{}^L \d_K{}^N - \fr15 \d_M{}^N \d_K{}^L.
\end{equation}
Upon the decomposition the $\bf{5}$ breaks into ${\bf 4}_1+{\bf 1}_{-4}$ where the subscript denotes weight w.r.t. the subalgebra $\frgl(1)$. Similarly the adjoint decomposes as  
\begin{equation}
    {\bf 24} \to {\bf 1}_0 + {\bf 4}_5 + \bar{{\bf 4}}_{-5} + {\bf 15}_0.
\end{equation}
More explicitly we have
\begin{equation}
    \begin{aligned}
         {\bf 5}: && V^M & \rightarrow (V^m,V^5), \\
         {\bf 24}: && T_M{}^N &\rightarrow (T_m{}^5, T_m{}^n, T_5{}^m).
    \end{aligned}
\end{equation}
For the deformation one takes $(T_{mnk})_M{}^N = \e_{mnkl}(T_5{}^l)_M{}^N$, that gives the following matrix
\begin{equation}
    \label{gl5_trans_matrix}
        O_M{}^N = 
        \begin{bmatrix}
            \delta_m{}^n & g^{\fr12}W^n \\
            0 & 1
        \end{bmatrix},\\
\end{equation}
where $W^m =  1/3!\ve^{mnkl}\W_{nkl}$. Denoting fields of the deformed background by capital letters $G_{\m\n}$, $\Phi$, $G_{mn}$, $V^m$ we obtain the following transformation rules
\begin{equation}
\label{Cdef}
\begin{aligned}
        K^{-1} &= 1 +   W_m W^m - 2 W_m v^m + \left(W_m v^m \right)^2,\\
        G_{\m\n} &= K^{-\frac{1}{3}}g_{\m\n},\\
        G_{m n} & = K^{\frac{2}{3}}\left( g_{mn} +(1+ v^2) W_m W_n - 2 v_{(m}W_{n)} \right), \\
     	C^{mnk} & = K^{-1}\Big(c^{mnk}+(1+ \frac{1}{3!}c^2) \W^{mnk}\Big).
\end{aligned}
\end{equation}
In the last line indices on $c_{mnk}$ are raised by the inverse undeformed metric $g^{mn}$, while indices of $C_{mnk}$ are raised by the inverse deformed metric $G^{mn}$. This set of transformations is a generalization of the map \eqref{eq:openclosed}.

To arrive at the generalized classical Yang-Baxter equation we take the tri-vector in the tri-Killing ansatz
\begin{equation}
    \W^{mnk} = \r^{abc}k_a{}^m k_b{}^n k_c{}^k,
\end{equation}
where $\r^{abc}$ is constant and completely antisymmetric. In order for the above transformation to give a solution to 11d supergravity equations it is sufficient to require
\begin{equation}    
    \label{eq:gCYBE}
    \begin{aligned}
        \rho^{a_1[a_2|a_6|}\rho^{a_3a_4|a_5|}f_{a_5a_6}{}^{|a_7]}-\rho^{a_2[a_1|a_6|}\rho^{a_3a_4|a_5|}f_{a_5a_6}{}^{|a_7]}=0,\\
        \rho^{a_1a_2a_3}f_{a_2a_3}{}^{a_4}=0.    
    \end{aligned}
\end{equation}
The first is usually referred to as generalized Yang-Baxter equation, while the second is referred to the unimodularity condition in analogy with a similar condition for bi-vector deformations. Relaxing the latter one arrives at a solution to equations of a generalization of 11d supergravity constructed in \cite{Bakhmatov:2022rjn,Bakhmatov:2022lin}. These are an 11d analogue of 10d generalized supergravity equations of \cite{Arutyunov:2015mqj}, that appear as a more general requirement for kappa-symmetry of the string. 

The above equation is equivalent to the vanishing R-flux condition found in \cite{Bakhmatov:2019dow} as a sufficient condition for such deformation to generate solutions. If the U-duality group is bigger than SL(5), say it is $\rmE_6$, the conditions have to be generalized to include additional 6-vector deformations \cite{Malek:2020hpo}
\begin{equation}
    \W^{m_1\dots m_6} = \r^{a_1 \dots a_6}k_{a_1}{}^{m_1}\dots k_{a_6}{}^{m_6}.
\end{equation}
In \cite{Gubarev:2020ydf} it has been shown that the conditions found in \cite{Malek:2020hpo} are sufficient for such polyvector deformations to generate solutions.

\section{Solutions to gCYBE for \texorpdfstring{$\mathfrak{so}(5)$}{}}

\label{sec:sols}

We are looking at solutions of the following equations with respect to $\r^{abc}$
\begin{equation}
\begin{aligned}
    \rho^{a_1a_2a_3}f_{a_2a_3}^{\ \ \ \, a_4}=0,&&\\
    \rho^{a_1[a_2|a_6|}\rho^{a_3a_4|a_5|}f_{a_5a_6}^{\ \ \ \, |a_7]}-\rho^{a_2[a_1|a_6|}\rho^{a_3a_4|a_5|}f_{a_5a_6}^{\ \ \ \, |a_7]}=0,
\end{aligned}
\label{the_general_eqations}
\end{equation}
where $a_1,a_2,\dots=1,\dots,N$ label isometries of the underlying manifold. We would be interested in both solutions of the whole system and in solutions to the first and second line separately. The first equation, that is the unimodularity condition, is linear and finding all its solutions for a given algebra is straightforward. We present all such solutions in Section \ref{sec:uni}. \

The second line in the case of interest is a set of 105 quadratic equations. Finding its general solution is computationally difficult. Hence, it is suggestive to investigate simple case to develop some intuition. For that let us start with the most simple compact non-abelian group, that is SO(3). Generators of the algebra $\frso(3)$ in Cartan basis satisfy the following commutation relations
\begin{equation}
\begin{aligned}
     &\left[H,E\right]=2E,&\\
       &\left[H,F\right]=-2F,&\\
       &\left[E,F\right]=H,&\\
\end{aligned}
\label{com__rel_so3}
\end{equation}
Structure constants $f_{ab}{}^c$ are then proportional to $\e^{abc}$ implying that the unimodularity condition cannot be satisfied. Hence, we learn that taking SO(3) as the isometry group one cannot construct unimodular deformations. This however does not mean that such deformations cannot generate solutions to 11d supergravity equations, as we show below. The same is true for the $\frso(4)$ algebra as it is simply a direct sum of two $\frso(3)$'s. Hence, the first algebra that can in principle render unimodular deformations is $\frso(5)$, which we are interested in here.

On the other hand deformations generated by $\frso(3)$ isometries trivially satisfy generalized Yang-Baxter equation simply because it involves antisymmetrization in four indices while we have only three generators. Hence, all such tri-vector deformations are Yang-Baxter. This is the key observation we will be using in what follows, as a large class of the solutions to \eqref{the_general_eqations} we find are build on $\frso(3)$ subalgebras of $\frso(5)$.

\subsection{Solutions in Cartan-Weyl basis}

\begin{figure}[H]

\centering

\begin{tikzpicture}
%\draw[step=0.5, very thin, gray] (0,-5) grid (12,0);
% \foreach \y in {0,...,12} \draw (\y,0.5) node{\y};

\draw (1,6) node (4) {$E_4$}; 
\draw (3,6) node (3) {$E_3$}; 
\draw (3,4) node (1) {$E_1$}; 
\draw (5,6) node (2){$E_2$}; 
\draw (5,4) node (0){$H_{1,2}$}; 
\draw (7,4) node (m1){$F_1$};
\draw (5,2) node (m2){$F_2$}; 
\draw (7,2) node (m3){$F_3$}; 
\draw (9,2) node (m4){$F_4$};

\draw[->] (4) edge[blue] (3) (3) edge[red] (1) (1) edge[blue] (0) (3) edge[blue] (2) (2) edge[red] (0) (0) edge[red] (m2) (0) edge[blue] (m1) (m1) edge[red] (m3) (m2) edge[blue] (m3) (m3) edge[blue] (m4);

\end{tikzpicture}
\caption{Weight diagram of the $\bf 10$ of $\mathfrak{so}(5)$. Blue arrows denote action of $F_1$, red arrows denote action of $F_2$.}
\label{fig:so5}
\end{figure}
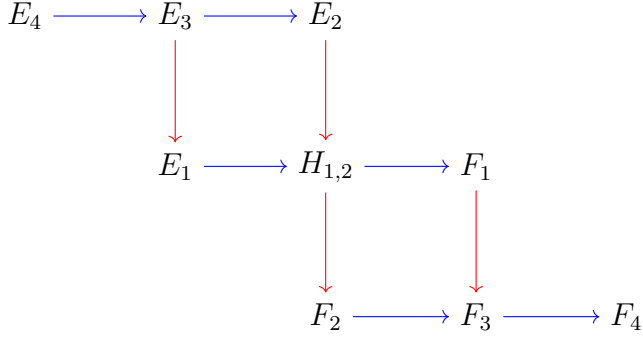

The compact group SO(5) is the isometry group of a 4-sphere $\SS^4$ and hence our analysis will cover tri-vector deformations of the AdS$_7\times \SS^4$ background. Generators $M_{AB}$ of the algebra $\frso(5)$ can be defined by the following commutation relations in terms of rotations in a five-dimensional flat space:
\begin{equation}
    \left[M_{AB},M_{CD}\right]=\eta_{BC}M_{AD}+\eta_{AD}M_{BC}-\eta_{AC}M_{BD}-\eta_{BD}M_{AC}
    \label{rot_gens_5}
\end{equation}
where here $A,B,C,D =1,\dots,5$ label the fundamental representation of $\frso(5)$ For our purposes it is convenient to turn to the orthogonal Cartan-Weyl basis where the generators are expressed in terms of $M_{AB}$ as follows 
\begin{equation}
\begin{aligned}
    H_{1}&= i M^{12},\    &H_{2}&= i M^{34},\\
    E_{1}&=\frac{1}{\sqrt{2}} \left(M^{45}+iM^{35}\right),\    &E_{2}&=\frac{1}{2} \left(M^{23}+iM^{13}-M^{14}+iM^{24}\right),\\
    E_{3}&=\frac{1}{\sqrt{2}} \left(M^{15}-iM^{25}\right),\    &E_{4}&=\frac{1}{2} \left(M^{23}+iM^{13}+M^{14}-iM^{24}\right),\\
    F_{1}&=\frac{1}{\sqrt{2}} \left(-M^{45}+iM^{35}\right),\    &F_{2}&=\frac{1}{2} \left(-M^{23}+iM^{13}+M^{14}+iM^{24}\right),\\
    F_{3}&=\frac{1}{\sqrt{2}} \left(-M^{15}-iM^{25}\right),\    &F_{4}&=\frac{1}{2} \left(-M^{23}+iM^{13}-M^{14}-iM^{24}\right).\\
\end{aligned}
\label{c_w_vars}\end{equation}
In these notations the generators $E_1$ and $E_2$ are the simple roots of the algebra with coordinates $(0,1)$ and $(1,-1)$ in the orthogonal basis respectively. Given $M_{AB}$ are real (Hermitean), negative root generators $F_i$ are related to positive root generators $E_i$ by conjugation:
\begin{equation}
    E_{\a}= -{\left(F_{\a}\right)}^{*},
\label{c_w_vars_rel}\end{equation}
where $\a,\b=1,\dots,4$.
The only non-vanishing commutation relations are the following
\begin{equation}
\begin{aligned}
    &\left[H_{1},E_{2}\right]=E_{2}, && \left[H_{1},F_{2}\right]=-F_{2}, &&\left[E_{1},E_{2}\right]=E_{3}, && \left[E_{1},F_{3}\right]=-F_{2},  &&\left[E_{3},F_{3}\right]=H_{1}\\
    &\left[H_{1},E_{3}\right]=E_{3}, && \left[H_{1},F_{3}\right]=-F_{3}, &&\left[E_{1},E_{3}\right]=E_{4}, && \left[E_{1},F_{4}\right]=-F_{3},  &&\left[E_{3},F_{4}\right]=F_1\\
    &\left[H_{1},E_{4}\right]=E_{4}, && \left[H_{1},F_{4}\right]=-F_{4}, && \left[F_{1},F_{2}\right]=-F_{3}, &&\left[E_{2},F_{2}\right]=H_{1}-H_{2},  &&\left[E_{4},F_{1}\right]=-E_{3}\\
    &\left[H_{2},E_{1}\right]=E_{1}, && \left[H_{2},F_{1}\right]=-F_{1}, && \left[F_{1},F_{3}\right]=-F_{4}, &&\left[E_{2},F_{3}\right]=F_{1},  &&\left[E_{4},F_{3}\right]=E_{1}\\
    &\left[H_{2},E_{2}\right]=-E_{2},&& \left[H_{2},F_{2}\right]=F_{2},  && \left[E_{1},F_{1}\right]=H_2, &&\left[E_{3},F_{1}\right]=-E_{2},  &&\left[E_{4},F_{4}\right]=H_{1}+H_{2}\\
    &\left[H_{2},E_{4}\right]=E_{4}, && \left[H_{2},F_{4}\right]=-F_{4},  &&\left[E_{3},F_{2}\right]=E_{1},  &&\\
\end{aligned}
\label{com__rel_so5}
\end{equation}
that can be nicely represented by Hasse diagram depicted on Fig. \ref{fig:so5}

To parameterize components of a tri-vector deformation tensor it is convenient  to label all the generators uniformly, for which purpose we introduce indices $a=\{\a,\bar{\a}, *, \bar{*}\}$ and define
\begin{equation}
\begin{aligned}
    T_{a}= \{T_\a,T_{\bar{\a}}, T_*,T_{\bar{*}}\} \equiv \{E_\a,F_\a,H_1,H_2\}.
\end{aligned}
\label{al_gen_so5}
\end{equation}
In these notations structure constants defined as usually as $[T_a,T_b] = f_{ab}{}^cT_c$ can be listed as follows
\begin{equation}
\begin{aligned}
    &f_{*\, 2}{}^2=1, && f_{*\, \bar{2}}{}^{\bar{2}}=-1, &&f_{1\, 2}{}^3=1, && f_{1\, \bar{3}}{}^{\bar{2}}=-1,  &&f_{3\,\bar{3}}{}^{*}=1\\
    &f_{*\, 3}{}^3=1, && f_{*\, \bar{3}}{}^{\bar{3}}=-1, &&f_{1\, 3}{}^4=1, && f_{1\, \bar{4}}{}^{\bar{3}}=-1,  &&f_{3 \bar{4}}{}^{\bar{1}}=1\\
    &f_{*\, 4}{}^4=1, && f_{*\, \bar{4}}{}^{\bar{4}}=-1, && f_{\bar{1} \, \bar{2}}{}^{\bar{3}}=-1 , &&f_{2\,\bar{2}}{}^{*}=1,  &&f_{4\, \bar{1}}{}^{3}=-1\\
    &f_{\bar{*}\, 1}{}^1=1, && f_{\bar{*}\, \bar{1}}{}^{\bar{1}}=-1, && f_{\bar{1}\, \bar{3}}{}^{\bar{4}}=-1, &&f_{2 \, \bar{3}}{}^{\bar{1}}=1,  &&f_{4 \, \bar{3}}{}^1=1\\
    &f_{\bar{*}\, 2}{}^2=-1,&& f_{\bar{*}\, \bar{2}}{}^{\bar{2}}=1,  && f_{1 \, \bar{1}}{}^{\bar{*}}=1, && f_{3\, \bar{1}}{}^2=-1,  &&f_{4\, \bar{4}}{}^*=1\\
    &f_{\bar{*}\, 4}{}^4=1, && f_{\bar{*}\, \bar{4}}{}^{\bar{4}}=-1,  && f_{3 \, \bar{2}}{}^1 = 1,  &&f_{2\,\bar{2}}{}^{\bar{*}}=-1, && f_{4\, \bar{4}}{}^{\bar{*}}=1
\end{aligned}
\label{com_rel_so5}
\end{equation}

In principle, both $\rho$'s with three and six indices can be turned on in the most general case. Our analysis using Wolfram Mathematica code shows that for the $\frso(5)$ algebra all solutions to gCYBE together with the unimodularity constraint generate complex $\W = \r^{abc}T_a\wedge T_b\wedge T_c$. Such deformation tensors are apparently of no interest as they generate complex solutions to supergravity equations.  Hence, we conclude that there are no real solutions to the set of equations \eqref{the_general_eqations} for the algebra $\frso(5)$. Closer inspection of the conditions gives a hint that such strong restriction comes from an interplay between the unimodularity constraint, linear in $\rho$'s, and the generalized Yang-Baxter equation, quadratic in $\r^{abc}$. Note however, that to generate solutions of 11d supergravity one is not necessarily interested exclusively in solutions to generalized Yang-Baxter equations together with the unimodularity constraint. First, as we discuss below and in more details in \cite{Gubarev:2020ydf} for SL(5) tri-vector deformations it is sufficient to impose only the unimodularity constraint. Such deformations are not Yang-Baxter. Second, to find Yang-Baxter tri-vector deformed solutions to 11d supergravity equations one may only satisfy the second line in \eqref{the_general_eqations} and to require 
\begin{equation}
    J^{mn} = \r^{abc}f_{bc}{}^d k_a{}^m k_d{}^n =0.
\end{equation}
In other words unimodularity constraint might be not satisfied, however give zero when contracted with Killing vectors. These two observations considerably simplify the problem and in this work we a heading at solutions of this type, leaving the most general analysis of the equations \eqref{the_general_eqations} for a future work. Hence, we are interested in i) solutions to the unimodularity constraint and ii) non-unimodular solutions to gCYBE. The former we analyze in Section \ref{sec:uni}, while here we focus at the latter.

It appears that the unimodularity condition imposes pretty strong constraints on $\r^{abc}$, such that solving gCYBE without it becomes a computationally involved task. Since we are interested in proving that gCYBE has non-trivial solutions along compact isometries it is natural to focus at finding of particular examples, leaving full analysis to a future work. For that we recall the observation  made in the beginning of this Section that $\frso(3)$ trivially solves gCYBE. Hence, we consider $\frso(3)$ subalgebras of $\frso(5)$. These appear to generate solutions to generalized Yang-Baxter equation, which are however non-unimodular. In addition, to  demonstrate that the space of solutions to generalized YB equation is not exhausted by taking $\frso(3)$ subalgebras we have found a different type of solution by introducing an object $s_{a,bc}$ that is inverse to $\r^{abc}$ in the sense of rectangular matrices. Altogether our solutions can be collected as follows
\begin{equation}
\label{omega_example}
\begin{aligned}
&1.&&\hat{\W}_1 =  a_1 \ E_2\wedge F_2\wedge \left(H_1-H_2\right)+a_2 \ E_4\wedge F_4\wedge \left(H_1+H_2\right)\\
&2.&&\hat{\W}_2 =  a \Big(\sqrt{3} E_1+\sqrt{2}\e E_2\Big)\wedge \Big(\sqrt{3} F_1+\sqrt{2}\e F_2\Big)\wedge \left(2H_1+H_2\right)\\
&3.&&\hat{\Omega}_3 =  a \ E_1\wedge F_1\wedge H_2\\
%&2.&&\hat{\Omega}_2 =  a \ E_2\wedge F_2\wedge \left(H_1-H_2\right)\\
%
&4.&&\hat{\Omega}_4 = a \ E_3\wedge F_3\wedge H_1\\
%
%&4.&&\hat{\Omega}_4 =  a \ E_4\wedge F_4\wedge \left(H_1+H_2\right)\\
%
&5.&&\hat{\Omega}_5 =  a  \big(  E_2+\e E_4\big)\wedge \big( F_2 + \e F_4\big)\wedge H_1\\
&6.&&\hat{\Omega}_6 =  a \Big[ E_1\wedge E_2\wedge F_3+
E_1\wedge E_3\wedge F_4-E_3\wedge F_1\wedge F_2-E_4\wedge F_1\wedge F_3 + \\ &&& \qquad +E_1\wedge F_1\wedge H_2+E_3\wedge F_3\wedge H_1+E_2\wedge F_2\wedge \left(H_1-H_2\right)+E_4\wedge F_4\wedge \left(H_1 + H_2\right)\Big],
\end{aligned}
\end{equation}
where  $\e= \pm1$. We are using hats to distinguish deformation written as wedge products of abstract generators from tri-vectors define as their geometric realizations in terms of Killing vectors. Note, that $\hat{\W}_1$ is a two-parametric solution, while in general a sum of two solutions to gCYBE is not a solution. The reason is that a procedure of generating solution to generalized YB equation analogous to that of Belavin-Drinfeld for the usual classical YB equation is not applicable, since it is not clear what general condition can be imposed for a sum of two solutions of gCYBE to be also a solution. All five solutions above are real, i.e. $\hat \W^* = \hat \W$, hence they generate real 11d backgrounds. However, there is a subtlety when trying to realize the above in terms of Killing vectors of the underlying space, to which we return momentarily.

The above results provide explicit examples that generalized Yang-Baxter equation in contrast to classical Yang-Baxter equation allows solutions that belong to $\frg\wedge \frg\wedge \frg$ where $\frg$ is algebra of a compact Lie group. We find that when $\frg=\frso(5)$ there are no real solutions to both unimodularity condition and generalized YB equation. The same is true for $\mathfrak{su}(3)$ where already unimodularity condition cannot be satisfied on itself. This suggests to consider larger and/or different compact groups e.g. $\frsu(4), \frso(6), \mathfrak{usp}(6)$, etc. Note however, that this does not mean that the obtained solutions cannot generate solutions to ordinary 11d supergravity equation. We will discuss this immediately.

\subsection{Geometric realization}

To consider the solutions to generalized Yang-Baxter equation found above as deformations of supergravity backgrounds the $\frso(5)$ generators must be realized as Killing vectors of the underlying space-time. For our applications here that include deformations of the internal 4-sphere  of the AdS$_7\times \SS^4$ and a and 3-sphere of $\RR^{1,10}$, the most convenient way to proceed is to start with rotation generators $M_{\mu\nu}$ of a  5-dimensional Euclidean plane:
\begin{equation}
M_{\hat\mu \hat\nu}=\frac{1}{2}\left(y_{\hat\mu}\partial_{\hat\nu}-y_{\hat\nu}\partial_{\hat\mu}\right),
\label{m_rep}
\end{equation}
where $\hat\mu,\hat\nu = 1,\dots,5$. Apparently these form the same $\frso(5)$ algebra as \eqref{rot_gens_5} and hence for the generators \eqref{c_w_vars} in the Cartan-Weyl basis 
we have
\begin{equation}
\begin{aligned}
    H_{1}&= \frac{i}{2}\big( y_{1} \partial_{2}-y_{2}\partial_{1}\big),\\
    H_{2}&= \frac{i}{2}\big( y_{3} \partial_{4}-y_{4}\partial_{3}\big),\\ 
    E_{1}&=\frac{i}{2\sqrt{2}}\Big[\big(y_3-iy_{4}\big) \partial_{5}- y_{5}\big(\dt_3-i\partial_{4}\big)\Big],\\
    E_{2}&=\frac{i}{4}\Big[\big(y_{1}-i y_{2}\big)\big(\dt_3 + i \dt_{4}\big)-\big(y_{3}+i y_{4}\big)\big(\dt_1 - i  \dt_{2}\big)\Big] ,\\
    E_{3}&=\frac{1}{2\sqrt{2}}\big[\big(y_{1}-i y_{2}\big) \partial_{5}-y_{5} \big( \partial_{1}-i \partial_{2} \big) \Big],\\    E_{4}&=\frac{i}{4}\Big[\big(y_{1}-i y_{2}\big)\big(\dt_3-i \partial_{4}\big)-\big(y_{3}-i y_{4}\big)\big( \dt_1-i\partial_{2}\big)\Big].
\end{aligned}
\label{c_w_vars_prevector}
\end{equation}
As before negative root generators $F_i$ are simply complex conjugates of $E_i$. It is worth to mention that we still consider the $\frso(5)$ algebra of Reals and the complex combinations are taken merely for convenience of further calculations. All final results are real and can be equivalently written in terms of $M_{\m\n}$ properly restricted to the submanifolds at question. 

Direct substitution of the above into \eqref{omega_example} gives that ${\Omega}_3={\Omega}_4={\Omega}_5={\Omega}_6=0$ are trivial, and hence only $\W_1$ and $\W_2$ can be used to generate tri-vector deformations. This follows from the following simple observations. First notice that the sets $\{E_1,F_1,H_2\}$ and $\{E_3,F_3,H_1\}$ form $\frso(3)$ subalgebras of $\frso(5)$. The same is true for $\{E_2,F_2, H_1-H_2\}$ and $\{E_4,F_4,H_1+H_2\}$, however for the former two sets we have
\begin{equation}
    \begin{aligned}
        E_1\wedge F_1 \propto \dt_5 \wedge H_2, \\
        E_3\wedge F_3 \propto \dt_5 \wedge H_1,
    \end{aligned}
\end{equation}
that is not true for the latter two.
Hence, multiplying the first by $H_2$ or the second by $H_1$ one gets zero. The reason is that the generators $\{E_1,F_1,H_2\}$ are realized as an algebra of rotational isometries of the 3D space parametrized by coordinates $\{y^3,y^4,y^5\}$. It is easy to see that in this case $E\wedge F\wedge H$ is always zero. In contrast the same is not true for the latter two sets of generators. The fact that there is no such a coordinate transformation can be seen in the so-called toroidal coordinates (see below), where the angle $\x$ corresponding to the action of the generator $H_1-H_2$ enters with an additional factor of $1/2$ in the generators $E_2$ and $F_2$.

\subsubsection{Isometries of \texorpdfstring{$\mathbb{S}_4$}{} }

Embedding of a 4-sphere of radius $r$ into the five-dimensional flat Euclidean space is given by the condition
\begin{equation}
    (y^1)^2+(y^2)^2+(y^3)^2+(y^4)^2+(y^5)^2=r^2.
\end{equation}
Let us turn to coordinates $\big(\r,\f,\s,\y,r\big)$ such that translation along the angles $\f$ and $\y$ correspond to the action of the Cartan operators $H_1$ and $H_2$ respectively. Explicitly the transition to these so-called toroidal coordinates on the 4-sphere has the following form
\begin{equation}
    \begin{aligned}
         y^1 & =  \r \cos \f, && y^3 & = \s \cos\y\\
         y^2 & = \r \sin \f, && y^4 & = \s \sin \y,\\
         (y^5)^2 & = r^2 - \r^2 - \s^2, 
    \end{aligned}
\label{doubled_spherical_cords}
\end{equation}
where $\f,\y \in [0,2\pi]$ and $\r^2+\s^2\leq r^2$. Fixing $r=R=$ const we obtain the metric on the $\SS^4$
\begin{equation}
    ds^2 = d\r^2 + \r^2 d\f^2 + d\s^2 + \s^2 d\y^2 + \fr{\big(\r d\r + \s d\s\big)^2}{R^2 - \r^2 - \s^2}.
\end{equation}
Similarly the rotational Killing vectors of the $\RR^5$ that form the group SO(5) can be written as follows
\begin{equation}
    \begin{aligned}
        k_* & = \fr i 2 \dt_\f, \quad k_{\bar{*}} = \fr i 2 \dt_ \y, \\
        k_1 &= -\fr{i}{2\sqrt{2}} e^{-i \y}\big(R^2 - \r^2 - \s^2\big) \big(\dt_\s - i \s^{-1}\dt_\y\big),\\
        k_2 & = -\fr i4 \s e^{-i(\f-\y)}\big(\dt_\r - i \r^{-1}\dt_\f\big)
        +\fr i4 \r e^{-i(\f-\y)}\big(\dt_\s + i \s^{-1}\dt_\y\big)\\
        k_3 &= -\fr{1}{2\sqrt{2}}e^{-i \f}\big(R^2 - \r^2 - \s^2\big)  \big(\dt_\r - i \r^{-1}\dt_\f\big),\\
        k_4 & = -\fr i4 \s e^{-i(\f+\y)}\big(\dt_\r - i \r^{-1}\dt_\f\big)
        +\fr i4 \r e^{-i(\f+\y)}\big(\dt_\s - i \s^{-1}\dt_\y\big). 
    \end{aligned}
\label{kill_vectors_sphere_doubled_cord}
\end{equation}
The tri-vector deformation tensor then takes the following simple form 
\begin{equation}
    \label{eq:omega_sph}
    \begin{aligned}
        &\W_1 = (a_1+a_2) \, \big(\r^2 + \s^2 \big)\Big(\fr{1}{\r} \dt_\r - \fr{1}{\s}\dt_\s\Big) \wedge \dt_\f \wedge \dt_\y.
    \end{aligned}
\end{equation}

As it has been mentioned in the previous section $\r^{abc}$ generating the above tri-vector are not unimodular, i.e. 
\begin{equation}
    \r^{acd}f_{cd}{}^d\neq 0.
\end{equation}
However, for a deformation to generate a supergravity solution it is enough that this expression contracted with Killing vectors vanishes. Explicitly one needs
\begin{equation}
    J^{mn}=\fr14 \r^{acd}f_{cd}{}^b k_a{}^m k_b{}^n=0.
\end{equation}
This condition is satisfied when $a_1=a_2=\k$, and in what follows we will always assume that. In this case the tri-vector deformations of supergravity backgrounds that we consider below are again solutions to the equations of 11d supergravity.

\subsubsection{Isometries of \texorpdfstring{$\mathbb{R}_4$}{}}

In what follows we will also be interested in tri-vector deformations of the flat 11d Minkowski space, or speaking more precisely in its 4D Euclidean part $\RR^4$. This space has compact SO(4) isometry whose embedding in the SO(5) isometry considered above is given by choosing $x^5=0$, i.e. we consider only rotations that keep the planes $(1,5),\dots,(4,5)$ fixed. Again it is convenient to choose toroidal coordinates on the 3-spheres of the $\RR^4$:
\begin{equation}
    \begin{aligned}
        y^1 & = \r \cos \f, && y^3  = \s \cos \y,\\
        y^2 & = \r \sin \f, &&  y^4  = \s \sin \y,
    \end{aligned}
    \label{tor_cor}
\end{equation}
with the radial distance given by $r^2 = \r^2 + \s^2$. Now $\f,\y \in [0,2\p]$ and $\r,\s > 0$. The metric takes the following form
\begin{equation}
    ds^2 = d\r^2 + \r^2 d\f^2 + d\s^2 + \s^2 d\y^2.
\end{equation}
Hence, we are dealing with only $k_*,k_{\bar{*}},k_2,k_4$ whose expressions in the chosen bases are given above. Apparently the tri-vector $\W_4$ has the same form
\begin{equation}
    \W_1 = 2(a_1+a_2)\big(\r^2 + \s^2\big)\Big(\fr1\r\dt_\r - \fr1\s\dt_\s\Big)\wedge \dt_\f \wedge \dt_\y,
\end{equation}
however with a different range of the coordinates $\r$ and $\s$, which now can be arbitrarily large. The non-unimodularity tensor $J^{mn}$ again vanishes for $a_1=a_2=\k$, that we will always assume in what follows.

\section{Examples}

\label{sec:ex}

Let us now use the results of the previous Section to tri-vector deform the particular solutions to 11d supergravity equations: the AdS$_{7}\times \SS^4$ background and the flat space-time $\RR^{1,10}$. Both backgrounds fit the truncation ansatz and hence the methods of \cite{Bakhmatov:2019dow,Bakhmatov:2020kul} briefly reviewed in Section \ref{sec:exft} can be immediately applied. Note, that in general the deformation does not respect all isometries of the initial background be it a 4-sphere or a flat manifold, that is expected. To determine the subalgebra of the preserved isometries it is suggestive to investigate which of the initial Killing vectors commute with the deformation, or in other words along which the Lie derivative of $\W^{mnk}$ vanishes. Explicit calculation gives:
\begin{equation}
\begin{aligned}
    &\mathcal{L}_{k_*}\W_1= \mathcal{L}_{k_{\overline{*}}}\W_1=0\\
    &\mathcal{L}_{k_2}\W_1=-i(a_1+a_2)e^{-i(\phi-\psi)} \, \big(\r^2 + \s^2 \big)\Big(\fr{\s}{\r^2} \dt_\r + \fr{\r}{\s^2}\dt_\s\Big) \wedge \dt_\f \wedge \dt_\y\\
    &\mathcal{L}_{k_4}\W_1=-i(a_1+a_2)e^{-i(\phi+\psi)} \, \big(\r^2 + \s^2 \big)\Big(\fr{\s}{\r^2} \dt_\r + \fr{\r}{\s^2}\dt_\s\Big) \wedge \dt_\f \wedge \dt_\y\\
    &\mathcal{L}_{k_1}\W_1=-e^{-i\psi}y_5(a_1+a_2) \, \big(\r^2 + \s^2 \big)\Big(-\fr{1}{\r} \dt_\r\wedge \dt_\s  + i\fr{1}{\s\r}\dt_\r\wedge \dt_\y- i\fr{1}{\s^2}\dt_\s\wedge \dt_\y \Big) \wedge \dt_\f\\
    &\mathcal{L}_{k_3}\W_1=-e^{-i\f}y_5(a_1+a_2) \, \big(\r^2 + \s^2 \big)\Big(- i\fr{1}{\s}\dt_\r\wedge \dt_\s  -\fr{1}{\s\r}\dt_\f\wedge \dt_\s-\fr{1}{\r^2} \dt_\r\wedge \dt_\f \Big) \wedge \dt_\y.
\end{aligned}
\end{equation}
We see, that at most the isometries along the Cartan generators $H_{1,2}$ can  be preserved. As we discuss below this maximally possible isometry group of the deformed background might be broken by a gauge choice of the 3-form and the actual isometry group becomes smaller.

\subsection{\texorpdfstring{AdS$_7\times \SS^4$}{AdS7xS4}}

Let us start with the AdS solution, that is of interest for holography applications. Explicitly the metric and the 4-form field strength have the following form
\begin{equation}
    \begin{aligned}
        d s^2 & = R^2 d s^2_{\SS^4} + 4 R^2 d s^2_{AdS_7}, \\
        F_4 & = \fr{3}{R} \text{Vol}_{\mathrm{S}_4}.  
    \end{aligned}
\end{equation}
Here $R$ is a constant that has the meaning of the radius of the AdS space and of the sphere. In the language of exceptional field theory the AdS space will be external while the sphere will be internal. Hence, the tri-vector deformation acts only along the coordinates on the sphere while the AdS part of the metric only acquires a prefactor. In the language of AdS/CFT correspondence such a deformation corresponds to an exactly marginal deformation of the dual gauge theory (leaving certain issues, such as stability, aside).
  
In the toroidal coordinates on the 4-sphere $(\r,\f,\s,\y)$ defined in \eqref{doubled_spherical_cords} the field strength takes the following form
\begin{equation}
    \begin{aligned}
        F = 3 \fr{R^5 \r\s}{\sqrt{R^2 - \r^2 - \s^2}} d\r \wedge d\f \wedge d\s \wedge d\y.
    \end{aligned}
\end{equation}
The particular form of its gauge potential $C_{mnk}$ depends on a gauge choice, and we find it convenient to set
\begin{equation}
    \begin{aligned}
        C_{\r\s\f } = \fr32 R^5\fr{\r\s}{\sqrt{R^2 - \r^2 - \s^2}}(\f+\y), \\
        C_{\r\s\y } = -\fr32 R^5\fr{\r\s}{\sqrt{R^2 - \r^2 - \s^2}}(\f+\y).
    \end{aligned}
\end{equation}
In this case components of the vector $v^m$ become simply
\begin{equation}
    v^\y =  \fr32 (\f+\y), \quad v^\f = \fr32 (\f+\y).
\end{equation}
One finds that in the chosen gauge the vector $v^m$ is orthogonal to the 1-form $W_{m}= 1/3! \e_{mnkl}\W^{nkl}$ that is 
\begin{equation}
    W_\r = \k R^{-1} \r \fr{\r^2 + \s^2}{\sqrt{R^2 - \r^2 - \s^2}}, \quad W_\s =\k R^{-1} \s \fr{\r^2 + \s^2}{\sqrt{R^2 - \r^2 - \s^2}}.
\end{equation}
Hence, given $W_m v^m\equiv 0$ the function $K$ takes the following simple form
\begin{equation}
    K^{-1}= 1+ \k^2 \fr{ (\r^2 + \s^2)^3}{R^6}
\end{equation}
Finally, we write the deformed background as follows
\begin{equation}
    \begin{aligned}
        ds^2 =&\ K^{\fr23}\Bigg[R^2ds_{\SS^4}^2  -\frac{3 R^6  \kappa  r_+^3 \psi _+ }{2\sqrt{1-r_+^2}} dr_+ \left(d\y_- r_-^2+d\y_+ r_+^2\right)
        +\frac{\kappa ^2 r_+^6 R^4 \left(9 R^4 r_+^2  \psi _+^2+4\right)}{4(1-r_+^2)}dr_+^2 \Bigg] \\
        &+ 4 R^2  K^{-\fr13}ds_{AdS_7}^2,\\
         F=&\Bigg[\frac{3 \left(2+4 \kappa ^2 r_+^6+27 \kappa ^2 r_+^8 R^4 \psi _+^2 \right)}{2 k^{5/6} \left(\kappa ^2 r_+^6+1\right)}\\ 
         &+\frac{3 \kappa  \sqrt{1-r_+^2} r_+^2 \left(4+2 \kappa ^2 r_+^6+3 r_+^2 R^4 \psi _+^2 \left(5 \kappa ^2 r_+^6+8\right)\right)}{2 k^{5/6} R^2 \left(\kappa ^2 r_+^6+1\right){}^2}\Bigg]d\r\wedge d\f\wedge d\s \wedge d\y
    \end{aligned}
\end{equation}
where we denote $r_\pm^2 = \r^2\pm\s^2$ and $\y_\pm= \f\pm\y$. Note that this is a solution to 11d supergravity equations as the tensor $J^{mn}=0$ and the generalized Yang-Baxter equation holds. In the language of exceptional field theory this means that generalized fluxes for this solution are precisely the same as those for the initial AdS${}_7\times \SS^4$ background.

We see that both the metric and the field strength do not depend on $\f-\y$  that indicates the remaining $\rmU(1)$ isometry corresponding to the action of $H_1 - H_2$. The other isometry commuting with $\W^{mnk}$ that is $H_1 + H_2$ is broken since the background explicitly depends on the corresponding angle coordinate $\y_+$. The reason for this comes from the gauge choice for the gauge potential we made above. Note first that Lie derivative of such chosen $C_{mnk}$ along $k_* + k_{\bar{*}}$ can be undone by a gauge transformation, hence $H_1+H_2$ is an isometry of the initial background as expected. This however is no longer true for the deformed background since the gauge potential enters the transformation non-linearly, and the transformation $\y_+ \to \y_+ + \chi(\y_+)$ cannot be undone by a gauge transformation. This manifests itself in the explicit dependence on $\y_+$. It is also evident that the background does not preserve any supersymmetry as it simply does not pass the test of \cite{Kulyabin:2022yls}. This is in consistency with the result of \cite{Cordova:2016xhm} stating that there are no supersymmetric marginal deformations of the $D=6$ $\mc{N}=(2,0)$ theory.

\subsection{Flat space-time}

To illustrate how gauge choice for the 3-form field affects the set of remaining isometries of the deformed background let us consider the case of an 11d flat space-time with vanishing 3-form. The metric is given by
\begin{equation}
d s^2 =d s^2 (\mathbb{R}^{4})+   d s^2 (\mathbb{R}^{1,6}), \quad
F_4 = 0,  
\end{equation}
with the obvious notations. As before the flat 7-dimensional space-time is considered to be the external space-time of the SL(5) exceptional field theory, while the 4d Euclidean space is understood as the internal. The deformation will be along the $\frso(4)$ subalgebra of $\frso(5)$ with the deformation tensor given by 
\begin{equation}
    W=\k \left(\rho ^2+\sigma ^2\right) ( \rho d\r + \s d\s ).
\end{equation}
Acting as before we obtain for the deformed background the following expressions
\begin{equation}
\begin{aligned}
    K^{-1}&=1+\k^2(\rho^2+\s^2)^3\\
    ds^2 & =K^{-\fr13}ds_{1,6}+K^{\fr23}\Big( ds_4+\k^2 \left(\rho ^2+\sigma ^2\right)^2 (\rho d\r + \s d\s)^2   \Big), \\
    F& =3 \k K^{-\fr56} \left(\rho ^2+\sigma ^2\right)\frac{ 2+\k^2 \big(\rho ^2+\sigma ^2\big)^3}{ \Big(1+\k^2 \big(\rho ^2+\sigma ^2\big)^3\Big)^2} d\r \wedge d\f \wedge d\s \wedge d\y.
\end{aligned}
\label{result_bg_R}
\end{equation}
This is again a solution to 11d supergravity equations since the generalized Yang-Baxter equation holds and the tensor $J^{mn}=0$. We see that in this case the whole $\rmU(1)\times \rmU(1)$ isometry generated by the Cartan subalgebra is preserved.

\subsection{Non-YB unimodular deformations}
\label{sec:uni}

Searching for tri-vector deformed supergravity backgrounds one recalls that in the SL(5) theory, i.e. when the internal manifold is four-dimensional, the  unimodular condition alone is sufficient for such a deformation to generate a supergravity solution. The same is true for bi-vector deformations ruled by the O(3,3) double field theory, i.e. when a deformation is given by Killing vectors of a 3d manifold. Interested in poly-vector deformations as solution generating transformations one certainly can drop the Yang-Baxter condition and search for all possible solutions to the unimodularity constraint. Hence, we are looking for all solutions to the following linear equations
\begin{equation}
    \r^{acd}f_{cd}{}^b=0,
\end{equation}
where $f_{ab}{}^c$ are structure constants of $\frso(5)$. We find in total 12 independent (complex) solutions, whose linear combinations again give a valid unimodular deformation. In terms of the $\frso(5)$ generators in the Cartan basis these read
\begin{equation}
\begin{aligned}
\hat{\Omega}^{1c}_{uni} &=  F_3\wedge F_4\wedge \left(H_2-H_1\right)+F_1\wedge F_2\wedge F_4 \\
\hat{\Omega}^{2c}_{uni} &=   E_1\wedge F_2\wedge \left(H_2+H_1\right)+E_4\wedge F_2\wedge F_3 \\
\hat{\Omega}^{3c}_{uni} &=  E_3\wedge E_4\wedge F_1-E_2\wedge E_4\wedge H_1+E_1\wedge E_2\wedge E_3   
\\
\hat{\Omega}^{4c}_{uni} &=  E_1\wedge F_3\wedge \left( H_1+H_2\right)+\left(E_1\wedge F_1+E_3\wedge F_3-E_4\wedge F_4\right)\wedge F_2+F_2\wedge H_1\wedge H_2 \\ 
\hat{\Omega}^{5c}_{uni} &= E_2\wedge E_3\wedge \left(H_1+H_2\right)-E_2\wedge E_4\wedge F_1\\
\hat{\Omega}^{6c}_{uni} &=   2 \left(E_3\wedge F_2+E_4\wedge F_3\right)\wedge H_1+E_1\wedge\left( E_2\wedge F_2-2 E_3\wedge F_3+ E_4\wedge F_4\right)+\\&+3 E_4\wedge F_1\wedge F_2-4 E_1\wedge H_1\wedge H_2 \\
\hat{\Omega}^{7c}_{uni} &= -E_3\wedge E_4\wedge F_2+E_1\wedge E_4\wedge \left(H_1-H_2\right) 
\\
\hat{\Omega}^{8c}_{uni} &=  (E_1\wedge E_2 +E_4\wedge F_1)\wedge F_3 \ +(E_4\wedge F_4-2E_1\wedge F_1 -E_2\wedge F_2)\wedge H_1 \\
&+( E_1\wedge F_4 - F_1\wedge F_2 ) \wedge E_3
\\
\hat{\Omega}^{9c}_{uni} &= ( E_1\wedge E_2 +E_4\wedge F_1)\wedge F_3 \ +(E_2\wedge F_2+E_4\wedge F_4-2E_3\wedge F_3)\wedge H_2\\
&+ (E_1\wedge F_4  -F_1\wedge F_2)\wedge E_3
\\
\hat{\Omega}^{10c}_{uni} &=    -E_4\wedge F_2\wedge H_2+E_1\wedge \left(E_3\wedge F_2-E_4\wedge F_3\right)
\\
\hat{\Omega}^{11c}_{uni} &=   \left(E_1\wedge F_1+E_2\wedge F_2-E_3\wedge F_3\right)\wedge E_4+E_1\wedge E_3\wedge H_1-E_1\wedge E_3\wedge H_2-E_4\wedge H_1\wedge H_2
\\
\hat{\Omega}^{12c}_{uni} &=-2 E_3\wedge F_2\wedge H_2+2 E_4\wedge F_3\wedge H_2+E_1\wedge E_2\wedge F_2-2 E_1\wedge E_3\wedge F_3\\
&+E_1\wedge E_4\wedge F_4+E_4\wedge F_1\wedge F_2
\\
\label{omega_unimod_example}
\end{aligned}
\end{equation}
Although one is apparently interested in real solutions we intentionally write the above as complex combinations of the generators. The reason is that since the unimodularity condition is a linear equation a linear combination of its solutions is again a solution. Hence, to end up with a valid unimodular non-Yang-Baxter real tri-vector deformation one simply takes a general combination of the above and of their complex conjugates. 

Instead of presenting explicitly the most general background resulting from a tri-vector deformation by a most general real linear combination of the above expressions, we write explicitly tri-vectors for each deformation on the 4-sphere as the most interesting case. To realize the above in terms of wedge products of the Killing vectors on the 4-sphere it is convenient to use semi-flat coordinates rather than the toroidal coordinates employed above. The reason is that to be expressed  in the most compact form each deformation requires its own choice of coordinates, while in semi-flat coordinates all the above can be written in an acceptable form. For the sphere $\SS^4$ the coordinates are defined as follows
\begin{equation}
    \begin{aligned}
        x^i &=y^i \text{ for } i=1,2,3,4, \\
        (y^5)^2 & = r^2 - x^i x^i.  
    \end{aligned}
\end{equation}
Then the 3-vector realizations of the above deformations become
\begin{equation}
\begin{aligned}
{\Omega}^{1c}_{\mathbb{S}uni}&=2 dx^{234} \left(x_1+i x_2\right) \left(x_4-i x_3\right) x_5-2 dx^{134} \left(x_1+i x_2\right) \left(x_3+i x_4\right) x_5-\\&-i dx^{124} \left(x_1+i x_2\right){}^2 x_5-dx^{123} \left(x_1+i x_2\right){}^2 x_5 ,\\
{\Omega}^{2c}_{\mathbb{S}uni}&=dx^{234} \left(-\left(x_3-i x_4\right){}^2\right) x_5+i dx^{134} \left(x_3-i x_4\right){}^2 x_5-\\&-2 dx^{124} \left(x_1+i x_2\right) \left(x_3-i x_4\right) x_5+2 dx^{123} \left(x_2-i x_1\right) \left(x_3-i x_4\right) x_5\\
{\Omega}^{3c}_{\mathbb{S}uni}&=dx^{234} \left(x_1-i x_2\right) \left(-2x_5^2+ x_1(x_1-i x_2)\right)+dx^{134} \left(x_2+i x_1\right) \left(-2x_5^2+ x_2(x_2+i x_1)\right)+\\&+dx^{124} \left(x_1-i x_2\right){}^2 x_3-dx^{123} \left(x_1-i x_2\right){}^2 x_4\\
{\Omega}^{4c}_{\mathbb{S}uni}&=dx^{234} \left(x_4+i x_3\right) \left(-x_5^2+\left(x_1+i x_2\right) x_1\right)+dx^{134} \left(x_3-i x_4\right) \left(-x_5^2+\left(x_2-i x_1\right) x_2\right)-\\&-dx^{124} \left(x_2-i x_1\right) \left(-x_5^2+\left(x_3-i x_4\right) x_3\right)-dx^{123} \left(x_1+i x_2\right) \left(-x_5^2+\left(x_4+i x_3\right) x_4\right)\\
{\Omega}^{5c}_{\mathbb{S}uni}&=2 dx^{234} \left(x_1-i x_2\right) \left(x_4-i x_3\right) x_5+2 dx^{134} \left(x_1-i x_2\right) \left(x_3+i x_4\right) x_5-\\&-i dx^{124} \left(x_1-i x_2\right){}^2 x_5-dx^{123} \left(x_1-i x_2\right){}^2 x_5\\
{\Omega}^{6c}_{\mathbb{S}uni}&=-4 i dx^{134} x_2 \left(x_3-i x_4\right) x_5-\frac{i}{4} dx^{124} x_5 \left(6 x_4\left( x_4+ix_3\right)-4 r^2+8 x_5^2- x_3^2-x_4^2\right)+\\&+\frac{1}{4} dx^{123} x_5 \left(-4 r^2+8 x_5^2- x_3^2-x_4^2+6x_3(x_3- i  x_4)\right)+4 dx^{234} x_1 \left(x_4+i x_3\right) x_5
\\
{\Omega}^{7c}_{\mathbb{S}uni}&=
dx^{234} \left(x_3-i x_4\right){}^2 x_5+i dx^{134} \left(x_3-i x_4\right){}^2 x_5+\\&+2 dx^{124} \left(x_1-i x_2\right) \left(x_3-i x_4\right) x_5+2 dx^{123} \left(x_1-i x_2\right) \left(x_4+i x_3\right) x_5\\
{\Omega}^{8c}_{\mathbb{S}uni}&=dx^{234} \left(-x_1\right) \left(x_1^2+x_2^2-4 x_5^2\right)+dx^{134} x_2 \left(x_1^2+x_2^2-4 x_5^2\right)-\\&-dx^{124} x_3 \left(x_1^2+x_2^2-2 x_5^2\right)+dx^{123} x_4 \left(x_1^2+x_2^2-2 x_5^2\right)\\
{\Omega}^{9c}_{\mathbb{S}uni}&=-dx^{234} x_1 \left(x_3^2+x_4^2-2 x_5^2\right)+dx^{134} x_2 \left(x_3^2+x_4^2-2 x_5^2\right)-\\&-dx^{124} x_3 \left(x_3^2+x_4^2-4 x_5^2\right)+dx^{123} x_4 \left(x_3^2+x_4^2-4 x_5^2\right)\\
{\Omega}^{10c}_{\mathbb{S}uni}&=dx^{234} x_1 \left(x_3-i x_4\right){}^2-dx^{134} x_2 \left(x_3-i x_4\right){}^2+\\&+dx^{124} \left(x_3-i x_4\right) \left(-2 x_5^2+x_3 \left(x_3-i x_4\right)\right)+dx^{123} \left(x_4+i x_3\right) \left(-2 x_5^2+x_4 \left(x_4+i x_3\right)\right)\\
{\Omega}^{11c}_{\mathbb{S}uni}&=dx^{234} \left(x_4+i x_3\right) \left(-x_5^2+x_1 \left(x_1-i x_2\right)\right)-dx^{134} \left(x_3-i x_4\right) \left(-x_5^2+x_2 \left(x_2+i x_1\right)\right)+\\&+dx^{124} \left(x_2+i x_1\right) \left(-x_5^2+x_3 \left(x_3-i x_4\right)\right)-dx^{123} \left(x_1-i x_2\right) \left(-x_5^2+x_4 \left(x_4+i x_3\right)\right)\\
{\Omega}^{12c}_{\mathbb{S}uni}&=
-8 i dx^{234} x_1 \left(x_3-i x_4\right) x_5-i dx^{124} x_5 \left(-4x_5^2+10 x_3(x_3- i  x_4)+x_3^2+ x_4^2\right)+\\&+dx^{123} x_5 \left(-4 x_5^2+10 x_4(x_4+ i x_3 )+ x_3^2+x_4^2\right)+8 dx^{134} x_2 \left(x_4+i x_3\right) x_5
\end{aligned}
\end{equation}
where we denote $dx^{ijk} = dx^i \wedge dx^j \wedge dx^k$ for compact notations. 

To conclude this section we note that the 3-vectors presented above generate in general a 24-parametric deformation of AdS${}_7\times \SS^4$, that preserves the AdS symmetry and satisfies equations of 11d supergravity. Interestingly enough these also include the deformation considered in the beginning of this Section for which explicit metric and field strength have been presented. Although this is generated by a non-unimodular $\hat{\W}_1$ satisfying generalized Yang-Baxter equations, the corresponding tensor $J^{mn}$ vanishes. We find that a combination of $\hat{\W}_{\SS uni}^{1c}$ and $\hat{\W}_{\SS uni}^{2c}$ generates precisely the same tri-vector $\W^{mnk}$ as $\hat{\W}_1$. Although it is tempting to state that the 24-parameter family of deformation found above exhausts all tri-vector deformations that generate a solution to 11d supergravity equations, this might not be true. Simply for the reason, that a non-unimodular $\hat{\W}$ might give vanishing $J^{mn}$. However, already in this form the results seem interesting for further investigation from the holographic point of view.

\section{Discussion}

\label{sec:conc}

In this work we consider tri-vector deformations of 11d supergravity backgrounds fitting the framework of the SL(5) exceptional field theory, i.e. transformations generated by $\W^{mnk} = \r^{abc}k_a\wedge k_b\wedge k_c$, where $k_a$ are Killing vectors of a four-dimensional submanifold and $\rho^{abc}$ is constant and totally antisymmetric. In particular we focus at the AdS$_7\times \SS^4$ and $\RR^{1,10}$ solutions to 11d supergravity equations. The Killing vectors are taken along non-commuting compact isometries of the 4-sphere and of a 3-sphere inside the flat space-time respectively. We find that generalized classical Yang-Baxter equation (gCYBE), that is a condition for such a deformation to generate a supergravity solution, has non-trivial solutions. This is in contrast to the standard classical Yang-Baxter equation relevant to bi-vector deformations generated by $\b=r^{ab}k_a\wedge k_b$, which has no solutions with non-commuting Killing vectors along compact isometries (see Appendix \ref{app:th}). 

We find explicit examples of such solutions to gCYBE and present explicit expressions for the deformed metric and 4-form field strength for the simplest of them. In addition we find a 24-parameter family of unimodular deformations that do not satisfy gCYBE, however still generate solutions to supergravity equations. This is a specific property of the SL(5) theory related to the dimension four of the manifold. All found deformations are non-abelian in the sense that they cannot be presented in the form $\W = k_1\wedge k_2\wedge k_3$ with all $k$'s commuting. Certainly, our solutions also cannot be presented in the form $\W = \b \wedge k$, where $k$ commutes with all generators inside $\b$. Existence of such solutions with non-abelian $\b$ is forbidden by the theorem in Appendix \ref{app:th}.

These results seem to be of interest in the context of AdS/CFT correspondence. Given the classification of \cite{Imeroni:2008cr} bi-vector deformations of AdS$_d \times M_{10-d}$ type solutions generated by Killing vectors of the compact manifold $M_{10-d}$ correspond to adding an exactly marginal operator to the dual CFT. The well-known example is the $\b$-deformation of \cite{Leigh:1995ep}. Similarly one may interpret tri-vector deformations along compact isometries: since the whole AdS isometry group is intact conformal symmetry of the dual theory does not change. Hence, it is natural to suggest that all such tri-vector deformations correspond to exactly marginal deformations. Our results show that in the tri-vector case one can go beyond simple abelian deformations. Certainly the issue of stability of the deformation has still to be investigated and further research is necessary to give precise interpretation of our solutions in terms of deformations of the $D=6$ $\mc{N}=(2,0)$ theory. For certain we known that all these are non-supersymmetric since the condition of \cite{Kulyabin:2022yls} does not hold. This is in consistency with \cite{Cordova:2016xhm} stating that there are no supersymmetric relevant or marginal deformations in $D=6$ SCFT's. It would be interesting to find gauge theory dual to our solutions in the six-dimensional theory.

Natural question is: whether our results can be applied to backgrounds dual to theories that are not IR free? More concretely: is it possible to find tri-vector non-abelian deformations along compact isometries of the AdS$_5\times \SS^5$ background? Since this is a solution to Type IIB supergravity equations the formalism of \cite{Bakhmatov:2019dow,Bakhmatov:2020kul} does not directly apply. However, on the other hand, exceptional field theory contains Type IIB supergravity and poly-vector deformations in principle survive the embedding. Moreover, it is natural to expect that these will be governed by precisely the same gCYBE as in the 11d case probably with additional terms on the RHS related to the SL(2) S-duality symmetry. Given it is true the solutions $\hat{\W}_1$ and $\hat{\W}_2$ we have found above will generate non-supersymmetric exactly marginal deformations of $D=4$ $\mc{N}=4$ SYM. Hence the question of defining poly-vector deformations in Type IIB theory is of most interest.

Another interesting question is: to what extent it is possible to extend the AdS/CFT correspondence beyond ordinary supergravity to cover backgrounds of generalized supergravity in 10d \cite{Arutyunov:2015mqj} or in 11d \cite{Bakhmatov:2022rjn,Bakhmatov:2022lin}? Such backgrounds are generated by non-unimodular bi- and tri-vector deformations respectively with non-vanishing $I^m$ or $J^{mn}$. To our knowledge there has been no particular progress in that direction. The reason might be simple: all bi-vector deformations along compact isometries are abelian and hence trivially unimodular. Bi-vector deformations along non-compact isometries that break AdS symmetries and generate non-commutative field theories can easily be made unimodular. Hence, in whole non-triviality this question arises when dealing with tri-vector deformations. All our examples are either unimodular or have vanishing $J^{mn}$ hence keep the deformed backgrounds in the set of supergravity solutions. 

Finally, the most straightforward extensions of the results presented here  are the following. First, more extensive analysis of generalized Yang-Baxter equations already for the algebra $\frso(5)$ is required. Although we have found that together with the unimodularity condition it has only complex solutions, it is worth to search for the most general set of solutions that give $J^{mn}=0$ and hence generate 11d supergravity solutions. Second, it would be interesting to apply the same approach to the AdS${}_4\times \SS^7$ background using the solutions we found here for $\frso(5)$ and also more general solutions for the full SO(8) isometry group. The results can be compared to the results of \cite{Ashmore:2018npi} where marginal deformations of the dual three-dimensional theory have been considered using generalized geometry.

\section*{Acknowledgments}

This work has been supported  by the Foundation for the Advancement of Theoretical Physics and Mathematics “BASIS”, grant No 21-1-2-3-1, by Russian Ministry of Education and Science and by the Russian Government Program of Competitive Growth of the Kazan Federal University.
\appendix

\section{Bi-vector deformations on compact isometries}
\label{app:th}

Classical Yang-Baxter equation takes the form:
\begin{equation}
    r^{i_1 k} r^{i_2 l } f_{k l }^{ \ \ i_3}+r^{i_3 k } r^{i_1 l}  f_{k l }^{ \ \ i_2}+r^{i_2 k } r^{i_3 l } f_{k l }^{ \ \ i_1}=0.
\label{cybe_eq}
\end{equation}
If we define inverse for $r$ matrix $s$, we can get next equation, by multiplying \eqref{cybe_eq} with $s_{i_1 \alpha} \ s_{i_2 \beta} \ s_{i_3 \gamma}$:
\begin{equation}
    f_{\alpha \beta }^{ \ \ i}s_{i \gamma}+  f_{ \gamma \alpha }^{ \ \ i }s_{i \beta}+f_{\beta \gamma }^{ \ \ i}s_{i \alpha}=0.
\label{2_cocycle_1}
\end{equation}
Also, by defining $s$ as two-form $B$ and remembering point of structure constant $f$, we can rewrite \eqref{2_cocycle_1} as 2-cocycle condition:
\begin{equation}
    B\left(\left[x,y\right],z\right)+B\left(\left[z,x\right],y\right)+B\left(\left[y,z\right],x\right)=0,
\label{2_cocycle_2}
\end{equation}
where $x,y,z$ - elements of subalgebra $L$ of some compact algebra $\mathfrak{g}$, on which we define \eqref{cybe_eq}. Without limiting the generality, we can thing,that $B(*,*)$ is non-degenerate form, which means that $\nexists \ a_{0} \in \mathfrak{g} : B(a_{0},x)=0, \ \forall x \in \mathfrak{g} $. \par
Next we can write as definition of non-abelian solution as condition $\forall x,y \in L, \ [x,y]\neq 0$. Let 's prove, that if $\mathfrak{g}$ is simple and compact, that $L$ is abelian. First, because $L$ ,obviously, is compact, we can use some special theorem, with which we can say, that L is direct sum of something semisimple derived algebra $L'$, and center of $L$ defined as $\xi \left(L\right)$:
\begin{equation}
    L=L'\oplus\xi \left(L\right)
\label{th1.1}
\end{equation}
Note, that derived algebra $L'$ is such subalgebra of $L$, that $L'\equiv [L,L]$, and $\xi \left(L\right) \in \left\{\forall x,y \in L|[x,y]=0 \right\}$. We can write \eqref{2_cocycle_2} with $\forall x,y \in L'$ and $\forall z \in \xi \left(L\right)$, which implies, by definition, $ B\left(\left[x,y\right],z\right)=0$, and because $x,y \in L'$, $	\exists \omega \in  L' : \omega=[x,y] $, then:
\begin{equation}
     B\left(\omega,z\right)=0, \forall z \in \xi \left(L\right),\forall \omega \in L'
\label{th1.2}
\end{equation}
On the other hand, since $L'$ is semisimple, the restriction of $B$ to $L'$ is a coboundary, i.e. there exists a non-zero
functional $f$ on $L'$ such that $B(w_1, w_2) = f ([w_1, w_2])$, $\forall w_1, w_2 \in L'$. Let $a_0$ be the element of $L'$ which
corresponds to $f$ via the isomorphism $L' \cong (L')^*$ deﬁned by the Killing form. Then $\forall w \in L'$ one has
\begin{equation}
     B(a_0, w) = K (a_0, [a_0, w]) = K ([a_0, a_0], w) = 0. 
\label{th1.3}
\end{equation}
Together with \eqref{th1.1} and \eqref{th1.2}, \eqref{th1.3} implies that
\begin{equation}
     B(a_0, l) =0,
\label{th1.4}
\end{equation}
$\forall l \in L$. Thus B is degenerate on L, which is a contradiction.

\bibliography{bib.bib}
\bibliographystyle{utphys.bst}

\end{document}